\documentclass[superscriptaddress,aps,preprintnumbers,amsmath,showpacs,amssymb,prd,nofootinbib,reprint,twocolumn,floatfix]{revtex4-1}
\usepackage{bm, color}
\usepackage{amssymb,amsfonts,slashed,amsthm,amsmath,graphicx,soul,empheq}
\usepackage[caption=false]{subfig}
\usepackage{hyperref}

\begin{document}

\newcommand{\vev}[1]{ \left\langle {#1} \right\rangle }
\newcommand{\bra}[1]{ \langle {#1} | }
\newcommand{\ket}[1]{ | {#1} \rangle }
\newcommand{\eV}{ \ {\rm eV} }
\newcommand{\KeV}{ \ {\rm keV} }
\newcommand{\MeV}{\  {\rm MeV} }
\newcommand{\GeV}{\  {\rm GeV} }
\newcommand{\TeV}{\  {\rm TeV} }
\newcommand{\1}{\mbox{1}\hspace{-0.25em}\mbox{l}}
\newcommand{\Red}[1]{{\color{red} {#1}}}

\newcommand{\lmk}{\left(}  
\newcommand{\rmk}{\right)}
\newcommand{\lkk}{\left[}  
\newcommand{\rkk}{\right]}
\newcommand{\lhk}{\left \{ }  
\newcommand{\rhk}{\right \} }
\newcommand{\del}{\partial}  
\newcommand{\la}{\left\langle} 
\newcommand{\ra}{\right\rangle}
\newcommand{\half}{\frac{1}{2}}

\newcommand{\bea}{\begin{array}}
\newcommand{\eea}{\end{array}}
\newcommand{\beq}{\begin{eqnarray}}
\newcommand{\eeq}{\end{eqnarray}}
\newcommand{\eq}[1]{Eq.~(\ref{#1})}

\newcommand{\dd}{\mathrm{d}}
\newcommand{\Mpl}{M_{\rm Pl}}
\newcommand{\mg}{m_{3/2}}
\newcommand{\abs}[1]{\left\vert {#1} \right\vert}
\newcommand{\mphi}{m_{\phi}}
\newcommand{\Hz}{\ {\rm Hz}}
\newcommand{\for}{\quad \text{for }}
\newcommand{\Min}{\text{Min}}
\newcommand{\Max}{\text{Max}}
\newcommand{\Kahler}{K\"{a}hler }
\newcommand{\cphi}{\varphi}
\newcommand{\Tr}{\text{Tr}}
\newcommand{\diag}{{\rm diag}}

\newcommand{\SUf}{SU(3)_{\rm f}}
\newcommand{\Upq}{U(1)_{\rm PQ}}
\newcommand{\Zpq}{Z^{\rm PQ}_3}
\newcommand{\Cpq}{C_{\rm PQ}}
\newcommand{\ubar}{u^c}
\newcommand{\dbar}{d^c}
\newcommand{\ebar}{e^c}
\newcommand{\nubar}{\nu^c}
\newcommand{\Ndw}{N_{\rm DW}}
\newcommand{\Fpq}{F_{\rm PQ}}
\newcommand{\fpq}{v_{\rm PQ}}
\newcommand{\Br}{{\rm Br}}
\newcommand{\Lag}{\mathcal{L}}
\newcommand{\Lqcd}{\Lambda_{\rm QCD}}

\newcommand{\ji}{j_{\rm inf}} 
\newcommand{\jb}{j_{B-L}} 
\newcommand{\M}{M} 
\newcommand{\im}{{\rm Im} }
\newcommand{\re}{{\rm Re} }

\def\lrf#1#2{ \left(\frac{#1}{#2}\right)}
\def\lrfp#1#2#3{ \left(\frac{#1}{#2} \right)^{#3}}
\def\lrp#1#2{\left( #1 \right)^{#2}}
\def\REF#1{Ref.~\cite{#1}}
\def\SEC#1{Sec.~\ref{#1}}
\def\FIG#1{Fig.~\ref{#1}}
\def\EQ#1{Eq.~(\ref{#1})}
\def\EQS#1{Eqs.~(\ref{#1})}
\def\TEV#1{10^{#1}{\rm\,TeV}}
\def\GEV#1{10^{#1}{\rm\,GeV}}
\def\MEV#1{10^{#1}{\rm\,MeV}}
\def\KEV#1{10^{#1}{\rm\,keV}}
\def\blue#1{\textcolor{blue}{#1}}
\def\red#1{\textcolor{blue}{#1}}

\newcommand{\eff}{\Delta N_{\rm eff}}
\newcommand{\neff}{\Delta N_{\rm eff}}
\newcommand{\cc}{\Omega_\Lambda}
\newcommand{\Mpc}{\ {\rm Mpc}}
\newcommand{\Msolar}{M_\odot}

\def\ft#1{\textcolor{blue}{#1}}
\def\FT#1{\textcolor{blue}{[{\bf FT:} #1]}}
\def\sn#1{\textcolor{red}{#1}}
\def\SN#1{\textcolor{red}{[{\bf SN:} #1]}}

\preprint{TU-1131}

\title{
Non-thermally trapped inflation by tachyonic dark photon production
}

\author{Naoya Kitajima}
\affiliation{Department of Physics, Tohoku University,Sendai, Miyagi 980-8578, Japan} 
\affiliation{Frontier Research Institute for Interdisciplinary Sciences, Tohoku University, Sendai, Miyagi, 980-8578 Japan}

\author{Shota Nakagawa}
\affiliation{Department of Physics, Tohoku University,Sendai, Miyagi 980-8578, Japan}

\author{Fuminobu Takahashi}
\affiliation{Department of Physics, Tohoku University,Sendai, Miyagi 980-8578, Japan}


\date{\today}

\begin{abstract}
We show that a dark Higgs field charged under U(1)$_{\rm H}$ gauge symmetry is trapped at the origin for a long time, if dark photons are produced by  an axion condensate via tachyonic preheating. The trapped dark Higgs can drive late-time inflation, producing a large amount of entropy. Unlike thermal inflation, the dark Higgs potential does not have to be very flat, because the effective mass for the dark Higgs is enhanced by large field values of dark photons with extremely low momentum.  After inflation, the dark Higgs decays into massive dark photons, which further decay into the SM particles through a kinetic mixing.  We show that a large portion of the viable parameter space is within the future experimental searches 
for the dark photon, because the kinetic mixing is bounded below for successful reheating. We also comment on the Schwinger effect which can hamper the tachyonic production of dark photons,
when the mass of dark photon is not the St\"{u}ckelberg 
mass, but is generated by the Higgs mechanism.
Such non-thermal trapped inflation could be applied to other cosmological scenarios such as the early dark energy, known as one of the solutions to the Hubble tension.
\end{abstract}


\maketitle

\flushbottom

\section{Introduction}
There generically exist moduli or axions in the superstring theories, and they acquire masses from supersymmetry (SUSY) breaking and/or non-perturbative effects.
If some of them remain light in the low-energy effective theory, they can have important impacts on cosmological evolution. For instance, if a light modulus field starts to oscillate about the potential minimum with an initial amplitude close to the Planck scale, it soon dominates the universe. Depending on the mass, it decays into the standard model (SM) particles at later times, which may spoil the success of the big bang nucleosynthesis (BBN) or exceed the observed cosmic-ray fluxes. Alternatively, if stable, the modulus field
may overclose the universe. This is known as the cosmological moduli problem~\cite{Coughlan:1983ci,Banks:1993en,deCarlos:1993wie}.\footnote{
Also, a modulus field generically decays into 
gravitinos~\cite{Endo:2006zj,Nakamura:2006uc,Dine:2006ii,Endo:2006tf}, other SUSY particles~\cite{Moroi:1999zb,Endo:2006zj,Nakamura:2006uc}, or its axionic partners~\cite{Cicoli:2012aq,Higaki:2012ar,Higaki:2013lra} with a large branching fraction, leading to similar cosmological problems.}

Various solutions to the cosmological moduli problem have been proposed~\cite{Yamamoto:1985rd,Lyth:1995ka,Dine:1998qr,Linde:1996cx,Randall:1994fr,Kawasaki:2004rx}. Among them, thermal inflation is low-scale inflation that lasts for at most a few tens of e-folds~\cite{Yamamoto:1985rd,Lyth:1995ka}. It produces a large amount of
entropy, diluting the moduli or axions which started to oscillate before thermal inflation. The entropy production required to solve the cosmological moduli problem was discussed in detail in Refs.~\cite{Asaka:1997rv,Asaka:1999xd}.

There are two important ingredients for thermal inflation to occur. One is the existence of thermal plasma, which keeps the flaton, the scalar field responsible for thermal inflation, at the origin for a while. The other is a very flat potential of the flaton field, as the name suggests.
Due to the flat potential,
the flaton develops a large vacuum expectation value (VEV) after thermal inflation.
The flatness of the potential can be ensured by a certain discrete symmetry in a SUSY set-up~\cite{Lyth:1995ka}.

In this paper, we propose a simple trapped inflation model that produces large entropy, using tachyonic production of dark photons from an axion condensate. The idea is to realize late-time inflation
like thermal inflation, by a dark Higgs field charged under U(1)$_{\rm H}$. Unlike thermal inflation, however, we consider dark photons whose momentum distribution is significantly deviated from the thermal one.
Dark photons are known to be produced from the axion condensate via tachyonic preheating, if the axion-dark photon coupling is sufficiently large~\cite{Garretson:1992vt}.
Such a large coupling of the axion to dark photons can be realized in a clockwork axion model~\cite{Higaki:2016yqk}.
The tachyonic production of photons or dark photons has been studied in literature in a variety of contexts; reduction of the QCD axion abundance~\cite{Agrawal:2017eqm,Kitajima:2017peg}, production of dark photon dark matter~\cite{Agrawal:2018vin,Co:2018lka,Bastero-Gil:2018uel}, generation of cosmological magnetic fields~\cite{Garretson:1992vt,Fujita:2015iga,Patel:2019isj}, emission of gravitational waves~\cite{Adshead:2018doq,Machado:2018nqk,Adshead:2019lbr,Adshead:2019igv,Machado:2019xuc,Ratzinger:2020koh,Namba:2020kij,Kitajima:2020rpm,Okano:2020uyr, Geller:2021obo}, etc.
As shown in Refs.~\cite{Kitajima:2017peg,Agrawal:2018vin}, the field value of the dark photon can be as large as the axion decay constant. Then, the effective mass for the dark Higgs will be of the same order. Such a non-thermal trapping is much more effective than using a thermal mass. As a result, the dark Higgs can drive inflation, even if the potential is a simple quartic Mexican-hat potential.
This should be contrasted to thermal inflation where the flaton potential must be extremely flat.
After the non-thermally trapped inflation ends, the dark Higgs develops a nonzero VEV and decays into massive dark photons. If the dark photons further decay into the SM particles through a kinetic mixing, a large entropy is produced, which can be used to solve the cosmological moduli problem. Given that the axion itself causes the cosmological moduli problem, it is interesting that in this scenario the axion triggers the entropy production that dilutes itself and other moduli.

In our scenario we will assume that the dark Higgs stays at the origin before the axion starts to oscillate. This is possible if the dark Higgs has a non-minimal coupling to gravity, and if the universe was dominated by oscillating inflaton or moduli fields. (In the case of radiation-dominated universe, one needs to have a large non-minimal coupling.)
The restoration of U(1)$_{\rm H}$ symmetry has important phenomenological advantages. First, dark photons are massless when the axion starts to oscillate. It implies that, even if it has a kinetic mixing with the SM photon (or a hypercharge gauge boson), a certain combination of them is totally decoupled from the SM particles. Thus, the tachyonic production of dark photons proceeds without any problems, since there are no light U(1)$_{\rm H}$-charged particles. If there were light charged particles, they would affect the evolution  of dark photons, and generically prevent the tachyonic growth. More physically, they would screen large electric fields.
Once the dark Higgs develops a VEV after inflation, dark photons become massive and the kinetic mixing with photons becomes physical.
Interestingly, the kinetic mixing is bounded below for successful reheating. It implies that such dark photons can be searched for at various experiments. The decay process of dark photons, which might be seen at experiments, is responsible for the reheating, i.e., the Big Bang.

The rest of this paper is organized as follows. In Sec.~\ref{sec:2} we provide the model of the dark Higgs, dark photon, and axion, and derive the relevant equations of motion. In Sec.~\ref{sec:inflation} we study the non-thermally trapped inflation, and also present numerical results of our lattice simulations. In Sec.~\ref{sec:4} we study if the produced entropy indeed solves the cosmological moduli problem. We discuss implications for dark photon search experiments in Sec.~\ref{sec:5}. The last section is devoted for discussion and conclusions.

\section{Set-up}
\label{sec:2}
We consider an Abelian Higgs model with an axion field. The Lagrangian is given by
\beq
\mathcal{L} &=& (D_\mu\Psi)^\dag D^\mu\Psi-V_\Psi(\Psi,\Psi^\dag)-\frac{1}{4}F_{\mu\nu}F^{\mu\nu}\nonumber\\
&+&\frac{1}{2}\del_\mu\phi\del^\mu\phi-V_\phi(\phi)-\frac{\beta}{4f_\phi}\phi F_{\mu\nu}\tilde{F}^{\mu\nu},
\label{Lagrangian}
\eeq
where $\Psi$ is the dark Higgs field with a charge $e$, $F_{\mu\nu}=\del_\mu A_\nu-\del_\nu A_\mu$ is the field strength tensor of a hidden U(1)$_{\rm H}$ gauge field $A_{\mu}$, $\tilde{F}^{\mu\nu}=\epsilon^{\mu\nu\rho\sigma}F_{\rho\sigma}/(2\sqrt{-g})$ is its dual with $g\equiv {\rm{det}}(g_{\mu\nu})$, 
and  $\phi$ is the axion. Here $D_\mu=\del_\mu-ieA_\mu$ is the covariant derivative, $f_\phi$ is the axion decay constant, and $\beta$ is the axion coupling with gauge bosons.
We will refer to the gauge boson $A_\mu$ as the  dark photon in the following.
We take the potentials of the dark Higgs and the axion as
\begin{align}
V_\Psi(\Psi,\Psi^\dag)&=\frac{\lambda}{4}\left(|\Psi|^2-v^2\right)^2
=V_0 - m_\Psi^2 |\Psi|^2 + \frac{\lambda}{4} |\Psi|^4,\\
V_\phi(\phi)&=m_\phi^2f_\phi^2\left[1-\cos\left(\frac{\phi}{f_\phi}\right)\right],
\end{align}
where $v$ is the VEV of the dark Higgs, $\lambda$ is the quartic coupling, $m_\phi$ is the axion mass, and we have defined $V_0 \equiv \lambda v^4/4$ and $m_\Psi^2 \equiv \lambda v^2/2$. For simplicity, we assume that the axion mass $m_\phi$ is constant with time
in the following analysis. 

The equations of motion can be derived from Eq.~(\ref{Lagrangian}) as 
\begin{align}
&\frac{1}{\sqrt{-g}}D_\mu(\sqrt{-g}D^\mu\Psi)+\frac{\del V_\Psi}{\del\Psi^*}=0,\\
&\frac{1}{\sqrt{-g}}\del_\mu(\sqrt{-g}\del^\mu\phi)+\frac{\del V_\phi}{\del\phi}+\frac{\beta}{4f_\phi}F_{\mu\nu}\tilde{F}^{\mu\nu}=0,\\
&\frac{1}{\sqrt{-g}}\del_\mu(\sqrt{-g}F^{\mu\nu})-2e{\rm{Im}}(\Psi^*D^\nu\Psi)+\frac{\beta}{f_\phi}\del_\mu\phi\tilde{F}^{\mu\nu}=0.
\end{align}
In the flat Friedmann-Lema$\hat{\text{\i}}$tre-Robertson-Walker (FLRW) universe, the above equations are reduced to
\begin{align}
&\ddot{\Psi}+3H\dot{\Psi}+\frac{\del V_\Psi}{\del\Psi^*}\nonumber\\
&-\frac{1}{a^2}\left(\nabla^2\Psi-2ieA_i\del_i\Psi-ie\Psi\del_iA_i-e^2A_iA_i\Psi\right)=0,\label{Higgseom}\\
&\ddot{\phi}+3H\dot{\phi}-\frac{1}{a^2}\nabla^2\phi+\frac{\del V_{\phi}}{\del\phi}+\frac{\beta}{f_\phi a^3}\epsilon_{ijk}\dot{A}_i\del_jA_k=0,\\
&\ddot{A}_i+H\dot{A}_i-\frac{1}{a^2}(\nabla^2A_i-\del_i\del_jA_j)-2e{\rm{Im}}(\Psi^*\del_i\Psi)\nonumber\\
&+2e^2|\Psi|^2A_i-\frac{\beta}{f_\phi a}\epsilon_{ijk}\left(\dot{\phi}\del_jA_k-\del_j\phi\dot{A}_k\right)=0,\label{HPeq}\\
&\del_i\dot{A}_i-2ea^2{\rm{Im}}(\Psi^*\dot{\Psi})-\frac{\beta}{f_\phi a}\epsilon_{ijk}(\del_i\phi)(\del_jA_k)=0,
\end{align}
where we adopt the temporal gauge $A_\mu=(0, A_i)$, and $A_\mu$ is the comoving field in the expanding universe, and the overdot represents the derivative with respect to time. Here
$a$ denotes the scale factor,  and we denote $\partial_i \partial^i = - a^{-2} \partial_i \partial_i = - a^{-2} \nabla^2$.
The last equation is the constraint equation since the longitudinal component of the dark photon is not dynamical. 

One can see that the second term from the end of the LHS of Eq.(\ref{Higgseom}) represents the effective mass squared for the dark Higgs, $e^2 {\bm A}^2$, induced by dark photons.\footnote{
Precisely speaking, the gauge-invariant expression for the effective mass squared of the dark Higgs is given by $|\partial_\mu \theta-e A_\mu|^2$, where $\theta \equiv {\rm arg}\left[\Psi\right]$, and $|\Psi| \ne 0$ is assumed.  We have numerically confirmed that the contribution of $\partial \theta$ is subdominant and
the effective mass is well approximated by $e^2 {\bm A}^2$ during the trapped regime in our gauge choice. 
} Here we define the spatially averaged value of the physical gauge field as $({\bm A})_i \equiv \sqrt{\langle A^2_i\rangle}/a$. We will see that, due to this effective positive mass squared, the dark Higgs field can be trapped at the origin for a long time.

\section{Non-thermally trapped inflation
\label{sec:inflation}}
\subsection{Initial condition\label{sec:setup}}
The non-thermally trapped inflation occurs as follows.
Let us denote the initial value of the axion as $\theta_* f_\phi$, where $\theta_*$ is the initial angle and its most natural value is ${\cal O}(1)$.
The axion  starts to oscillate about the potential minimum when the Hubble parameter becomes comparable to the axion mass, $H \sim m_\phi$.
Then, dark photons with relatively low momenta are efficiently produced from the axion condensate via tachyonic preheating. As a result, the dark Higgs field is trapped at the origin due to the large effective mass induced by dark photons, and it drives late-time inflation to produce a large amount of entropy. 

In the above scenario,
we assume that the dark Higgs field stays at the origin and the U(1)$_{\rm H}$ symmetry remains unbroken until the axion starts to oscillate. To this end we introduce a non-minimal coupling to gravity,
\beq
\mathcal{L} \supset -\xi R|\Psi|^2,
\label{Ricci}
\eeq
where $\xi$ is a numerical coefficient. The most natural values of $\xi$ is of order unity.
In the flat FLRW universe,
the Ricci curvature is given by $R =  6(\ddot{a}/a+(\dot{a}/a)^2)$. 
In the following we assume that the universe is
in the matter dominated phase where  either the inflaton or moduli fields is the major component. 
In the matter-dominated universe, we have $R = 3 H^2$, and the dark Higgs acquires an effective mass of order the Hubble parameter. The effective potential around the origin is given by
\beq
V_\Psi^{(\rm eff)}(\Psi,\Psi^\dag) = V_0 + (3\xi H^2 - m_\Psi^2)|\Psi|^2 + \cdots.
\eeq
Thus, if $m_\Psi^2 \lesssim 3 \xi m_\phi^2$,
the dark Higgs stays at the origin
until the onset of the axion oscillations. 
We will come back to this condition 
and discuss the case in which the dark Higgs initially develops a nonzero VEV
in Sec.~\ref{sec:conclusion}. On the other hand, if the universe is radiation dominated, the Ricci curvature vanishes at the classical level due to the conformal symmetry, but at the quantum level it is of  ${\cal O}(10^{-2}) H^2$. Thus, one may stabilize the Higgs at the origin in this case by taking $\xi$ to be of ${\cal O}(10^2)$.\footnote{Alternatively, one may assume that a small amount of dark photons and/or Higgs are produced by the inflaton decay, but the hidden sector is decoupled from the SM. In this case, the dark Higgs acquires a (tiny) thermal mass.}

The fact that the dark Higgs field is at the origin and the U(1)$_{\rm H}$ symmetry is restored is very important for the tachyonic production of dark photons, which we will discuss in the next subsection. For one thing, tachyonic preheating occurs efficiently because the dark photon is massless; if the dark photon is heavier than the axion, the tachyonic production is suppressed~\cite{Agrawal:2018vin}.
Furthermore, because the dark photon is massless, even if U(1)$_{\rm H}$ has a kinetic mixing with hypercharge, a certain combination of the gauge bosons is completely decoupled from the SM particle, and its  tachyonic production is not disturbed by the charged particles. These issues were discussed in detail in a similar  production
mechanism for massive dark photon dark matter~\cite{Agrawal:2018vin}.

\subsection{Tachyonic production of dark photons}
The axion is considered to be nearly homogeneous for a while after the onset of oscillations.
Then, the equation of motion for the dark photon is simplified as
\beq \label{eq:Apm}
\ddot{A}_{\bm{k},\pm}+H\dot{A}_{\bm{k},\pm}+\frac{k}{a}\left(\frac{k}{a}\mp\frac{\beta\dot{\phi}}{f_\phi}\right)A_{\bm{k},\pm}=0,
\eeq
where $A_{\bm{k},\pm}$ is the Fourier component of the dark photon field in the circular polarization basis. 
For $k/a < \beta|\dot\phi|/f_\phi$, either of the two circular polarization modes grows exponentially, depending on the sign of $\dot\phi$~\cite{Garretson:1992vt}.
The tachyonic growth of the dark photon is so efficient that the energy density of the dark photon soon becomes comparable to that of the axion. 
After that, the dynamics enters the nonlinear regime and the linear analysis is no longer applicable~\cite{Kitajima:2017peg}. We need  numerical lattice simulations  to follow the subsequent evolution.

Let us estimate when the energy density of dark photons becomes comparable to that of the axion and the system enters the non-linear regime. 
One can see that the dominant growing mode is $k_{\rm{peak}}/a \sim \beta |\dot\phi|/(2f_\phi) \sim \beta m_\phi |\phi|/(2f_\phi)$
from Eq.~(\ref{eq:Apm}), where $|\phi|$ denotes the oscillation amplitude. It takes the maximal value, $k_{\rm{peak}}/a \sim \beta m_\phi\theta_*/2$, at the onset of the axion oscillations, and it gradually decreases proportional to the oscillation amplitude. 
For the efficient tachyonic production of dark photons, we need $\beta \theta_* = {\cal O}(10)$.
The system enters the non-linear regime
when
\beq
\frac{1}{2}m_\phi^2f_\phi^2\theta_*^2\left(\frac{a_{\rm{osc}}}{a_{\rm nl}}\right)^3\simeq \frac{k_{\rm{peak}}^2(t_{\rm{nl}})}{2a_{\rm nl}^2}|{\bm{A}_{\rm nl}}|^2,
\label{nl}
\eeq
where we have
approximated the energy of dark photons to the gradient energy of the dominant growing mode,
and variables evaluated at the beginning of the oscillation are labeled with `osc', and those evaluated when entering the non-linear regime are labeled with `nl'.
Thus, the field value of the dark photon can be estimated as
\beq \label{eq:Anl}
|\bm{A}_{\rm nl}| \simeq \frac{2f_\phi}{\beta},
\eeq
which is typically a few orders of magnitude smaller than 
the axion decay constant. Note that such a large field value 
is due to the non-thermal production of dark photons with low momenta.
As a result, the dark Higgs acquires a very heavy mass as one can see from Eq.~(\ref{Higgseom}).

From the above linear equation of motion, we obtain the exponential growth factor of the mode with the wave number $k_{\rm{peak}}/a$ as
\begin{align}
&\exp\left(\frac{1}{2}\int^{t_{\rm nl}}_{t_{\rm osc}} dt\frac{\beta m_\phi|\phi(t)|}{2f_\phi}\right)\nonumber\\
&\simeq \exp\left[\frac{\beta m_\phi\phi_{*}}{2\pi f_\phi} \int^{t_{\rm nl}}_{t_{\rm osc}} dt \left(\frac{a(t)}{a_{\rm{osc}}}\right)^{-\frac{3}{2}} \right] \simeq \left(\frac{a_{\rm{nl}}}{a_{\rm{osc}}}\right)^{\frac{\beta\theta_*}{2\pi}},
\label{eq:growth}
\end{align}
where the oscillation part of $\phi(t)$ is integrated out as the averaged value,
and we used the relation, $H_{\rm osc} = 2/(3 t_{\rm osc}) \simeq m_\phi$.
Note that $\frac{1}{2}$ in the first term means that the enhancement of each helicity mode is switched every half a period. The initial field value of dark photon is roughly given by $|\bm{A}_{\rm{osc}}| \sim k_{\rm{peak}}/a_{\rm{osc}}$, and thus, from Eqs.~(\ref{eq:Anl}) and (\ref{eq:growth}), we obtain
\beq \label{eq:anl-aosc}
\frac{a_{\rm nl}}{a_{\rm osc}} \sim \left(\frac{4f_\phi}{\beta^2m_\phi\theta_*}\right)^{\frac{2\pi}{\beta\theta_*}}.
\eeq
Note that since the dominant growing mode changes with time due to the cosmic expansion, one needs numerical simulation to calculate precisely when the non-linear regime begins.
However, the above estimation shows a good agreement with the numerical calculation within a factor of $O(1)$, partly because of the very rapid growth of the instabilities.

\subsection{Duration of non-thermally trapped inflation
\label{sec:duration}}
Once the dark Higgs acquires a very large mass, it remains trapped at the origin until the field value of the dark photon becomes small enough. 
Here let us estimate the number of e-folds of the trapped inflation. 

The number of e-folds is defined by 
\begin{align}
    {\cal N } = \ln\left(\frac{a_{\rm end}}{a_{\rm begin}} \right),
\end{align}
where the subscripts ``begin" and ``end" mean that the variables are evaluated when the non-thermally trapped inflation begins and ends, respectively. 
Let us further decompose the ratio of the scale factors as
\begin{align}
    \frac{a_{\rm end}}{a_{\rm begin}} 
    = \left(\frac{a_{\rm osc}}{a_{\rm begin}}\right)
    \left(\frac{a_{\rm end}}{a_{\rm nl}}\right)
     \left(\frac{a_{\rm nl}}{a_{\rm osc}} \right).
\end{align}
In the following, we will evaluate these ratios of the scale factors in turn.

The non-thermally trapped inflation begins when the vacuum energy of the dark Higgs potential comes to dominate the universe. Since the universe was matter-dominated, we have
\beq
\frac{a_{\rm{osc}}}{a_{\rm{begin}}}\simeq
\left(\frac{V_0}{3 m^2_\phi \Mpl^2}\right)^{1/3}
=
\left(\frac{\lambda v^4}{12m^2_\phi \Mpl^2}\right)^{1/3},
\label{eq-start}
\eeq
where we have used the fact that the axion starts to oscillate when $H\sim m_\phi$.
On the other hand, the inflation ends when the dark Higgs field is destabilized from the origin. Since the dark Higgs field is trapped by the effective mass due to the large field value of dark photon, the destabilization occurs when the spatially averaged value of the effective mass $\langle e|{\bm{A}}| \rangle$ becomes smaller than the tachyonic mass $m_\Psi = \sqrt{\lambda/2}v$.
Since the physical value of the dark photon field decreases inversely proportional to the scale factor after the end of the tachyonic production (i.e. $a>a_{\rm nl}$), the scale factor at the end of inflation can be calculated by using Eq.~(\ref{eq:Anl}) as follows,
\beq
\frac{a_{\rm{end}}}{a_{\rm nl}}\simeq
e \frac{2 f_\phi}{\beta}
\frac{1}{m_\Psi}.
\label{nl-end}
\eeq
Thus, combining (\ref{eq:anl-aosc}),(\ref{eq-start}), and (\ref{nl-end}), we obtain
\beq
\mathcal{N}
&\simeq& 10 + \kappa
\left(37-\log (\beta^2\theta_*) \right) 
+\log\left(\frac{e}{\beta \lambda^\frac{1}{3}}\right) + 
\frac{1}{3} \log\left(\frac{m_\Psi}{0.1 m_\phi}\right) \nonumber \\ &&
+ 
\left(1+\kappa\right) \log\left(\frac{f_\phi}{0.1 \Mpl}\right)
- \left(\frac{1}{3}+\kappa\right) \log\left(\frac{m_\phi}{10^2{\rm GeV}}\right)
\label{efold}
\eeq
with 
\beq
\kappa \equiv \frac{2\pi}{\beta\theta_*} \simeq 0.2\, \theta_*^{-1} \left(\frac{30}{\beta}\right).
\eeq
Thus, the number of e-folds is typically about $10 - 20$. For instance, we have the e-folding number $\mathcal{N}\simeq12$ for $f_\phi=10^{17}\GeV$ $m_\phi=10^2\GeV$, $m_\Psi=10\GeV$, $\beta=30$, and $\lambda=e=1$.
Note that the duration of the inflation is mainly determined by the hierarchy between $f_\phi$ and $m_\phi$, i.e. the field value and the momentum of the dark photon.

\subsection{Results of numerical lattice calculations
\label{sec:numerical}}
To confirm the above analytic estimate, we have performed numerical lattice simulations to follow the dynamics of the axion, dark photon and Higgs fields.
Due to the exponential growth of dark photons, the dynamics of this system becomes highly nonlinear, requiring lattice simulations.
In our simulation, we set the grid number $N_{\rm grid} = 512^3$, the comoving box size $L_{\rm box} = 0.5 \pi m_\phi^{-1}$, and the initial conformal time $\tau_i = 0.1/m_\phi$ corresponding to $H=20m_\phi$.
We set the parameters as follows: $f_\phi=5\times 10^{17}$ GeV, $m_\phi=5 \times 10^8\GeV$, $v=f_\phi$, $\beta=30$, $e=2m_\phi/f_\phi$, $\lambda=10^{-4}m_\phi^2/v^2$ and $\xi=1/3$.
The initial values of the axion and the dark Higgs are $\theta_*=1$ and $|\Psi|/v=0.01$ respectively. 
The initial value of the dark photon is given by the vacuum fluctuation.

\begin{figure*}[tp]
\centering
\includegraphics[width=8.5cm]{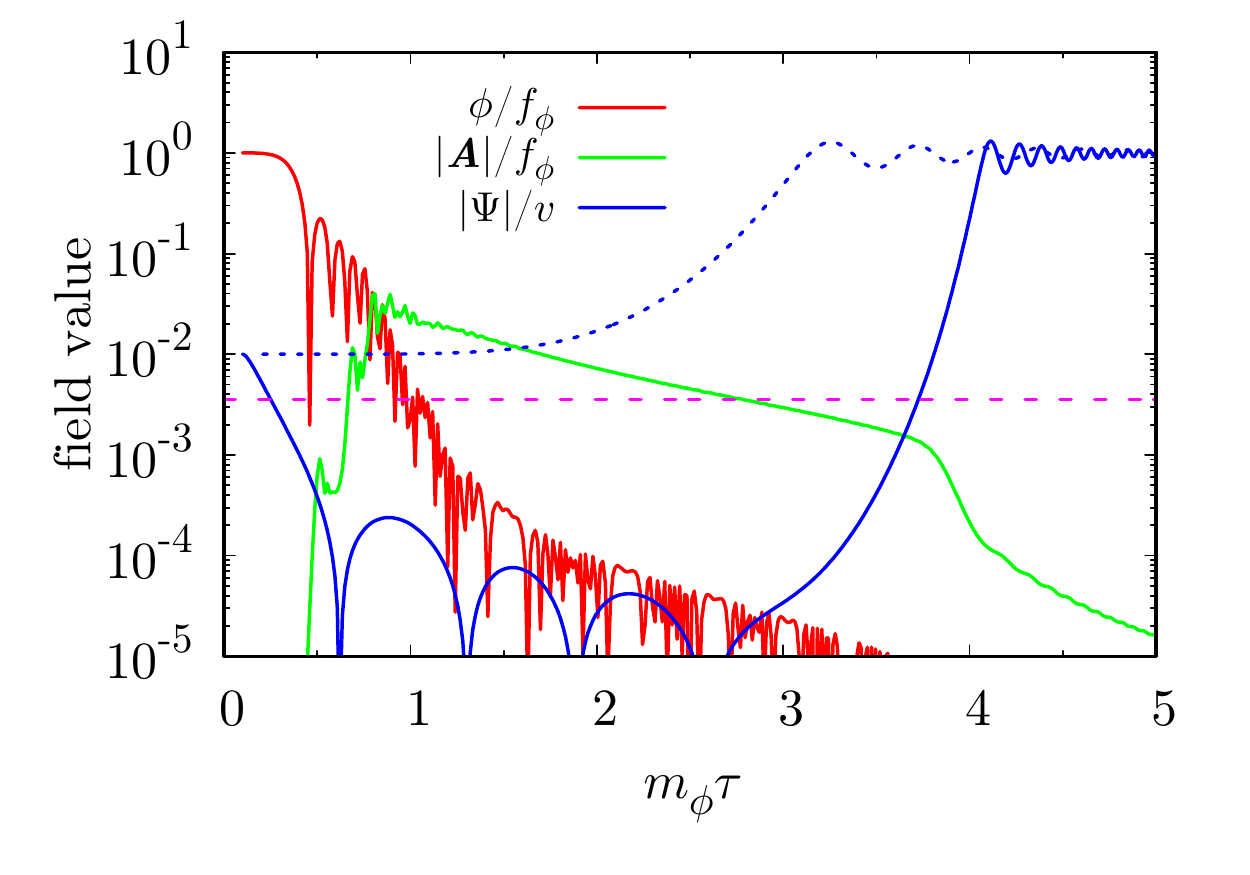}
\includegraphics[width=8.5cm]{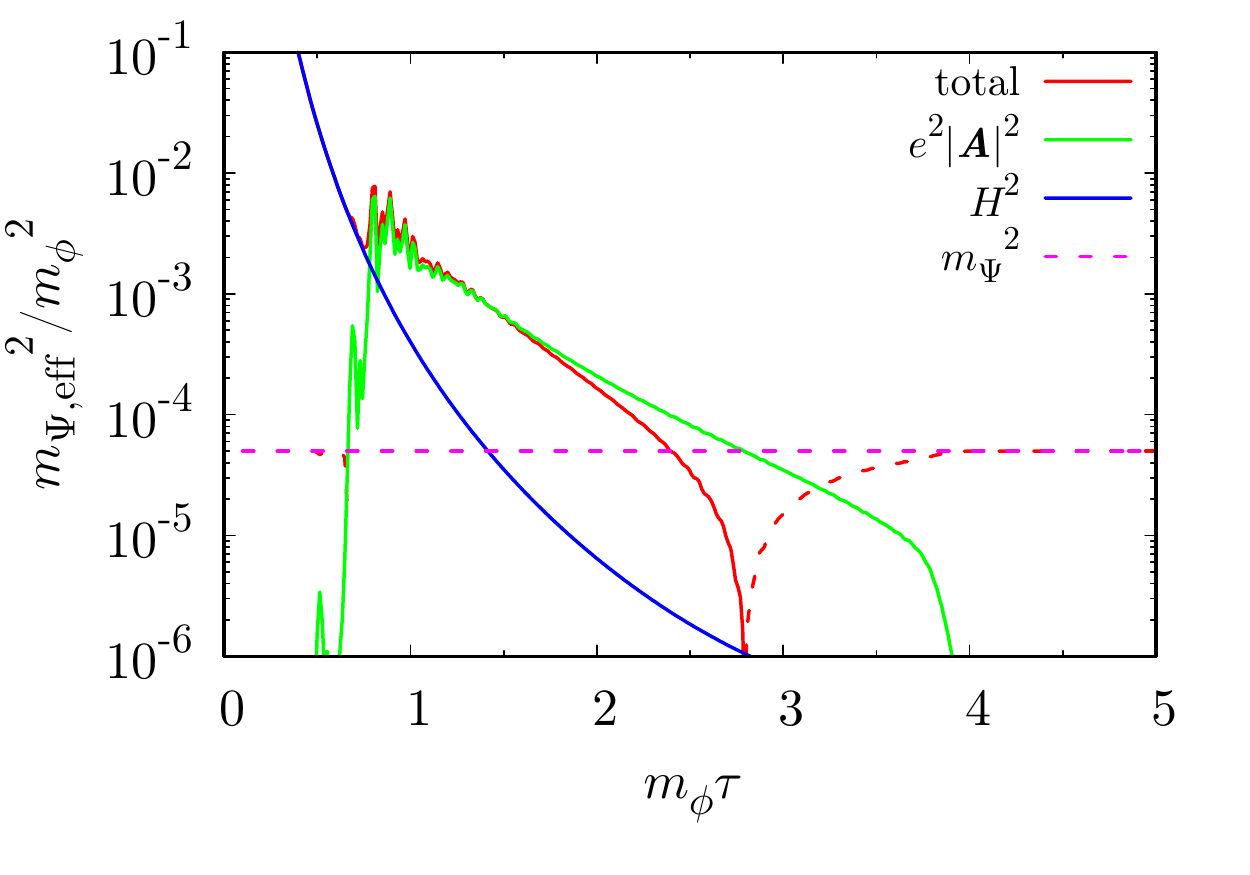}
\caption{
Left : time evolution of the spatial averaged field values of the axion (red), dark photon (green), both of which are normalized by $f_\phi$, and dark Higgs field (blue) normalized by $v$. The horizontal axis is the conformal time normalized by the axion mass. The dotted line corresponds to the case without dark photon production, i.e. $\beta=0$. The horizontal dashed line corresponds to $m_\Psi/(ef_\phi)$, 
so the intersection between the green line and the horizontal dashed line corresponds to $ e |{\bm A}|  = m_\Psi$.
Right : Time evolution of the effective dark Higgs mass squared. The red line shows the total and the contributions from the dark photon field, the Hubble-induced mass and the bare mass are shown by the red, green and blue lines respectively. The solid and dashed line corresponds respectively to positive and negative value. For $m_\phi \tau \gtrsim 3$, the effective mass at the origin is dominated by the (tachyonic) bare mass.
We set the parameters as follows: $f_\phi=5\times 10^{17}$ GeV, $m_\phi=5\times 10^8\GeV$, $v=f_\phi$, $\beta=30$, $e=2m_\phi/f_\phi$ and $\lambda=10^{-4}m_\phi^2 /v^2$.}
\label{fig:field_higgsmass}
\end{figure*}

\begin{figure*}[tp]
\centering
\includegraphics[width=8.5cm]{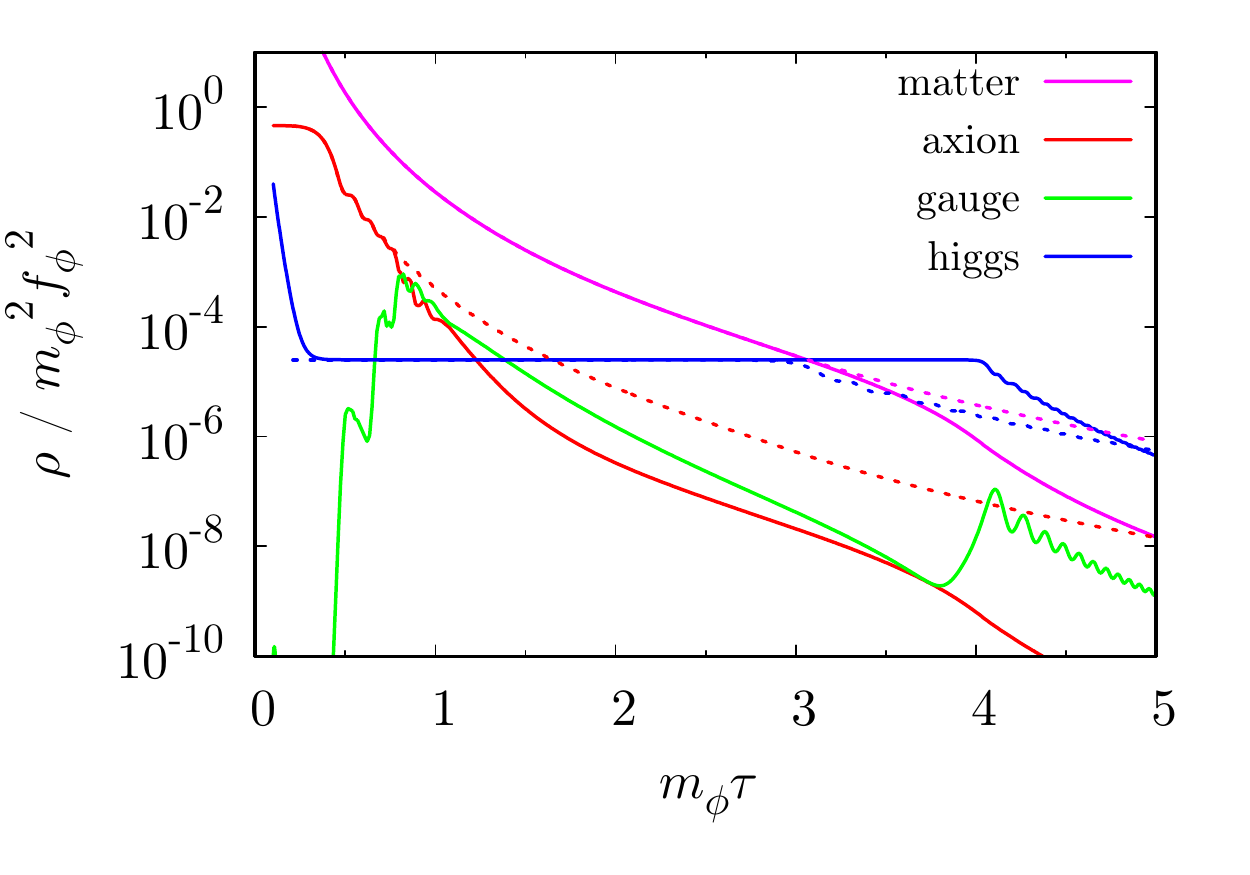}
\includegraphics[width=8.5cm]{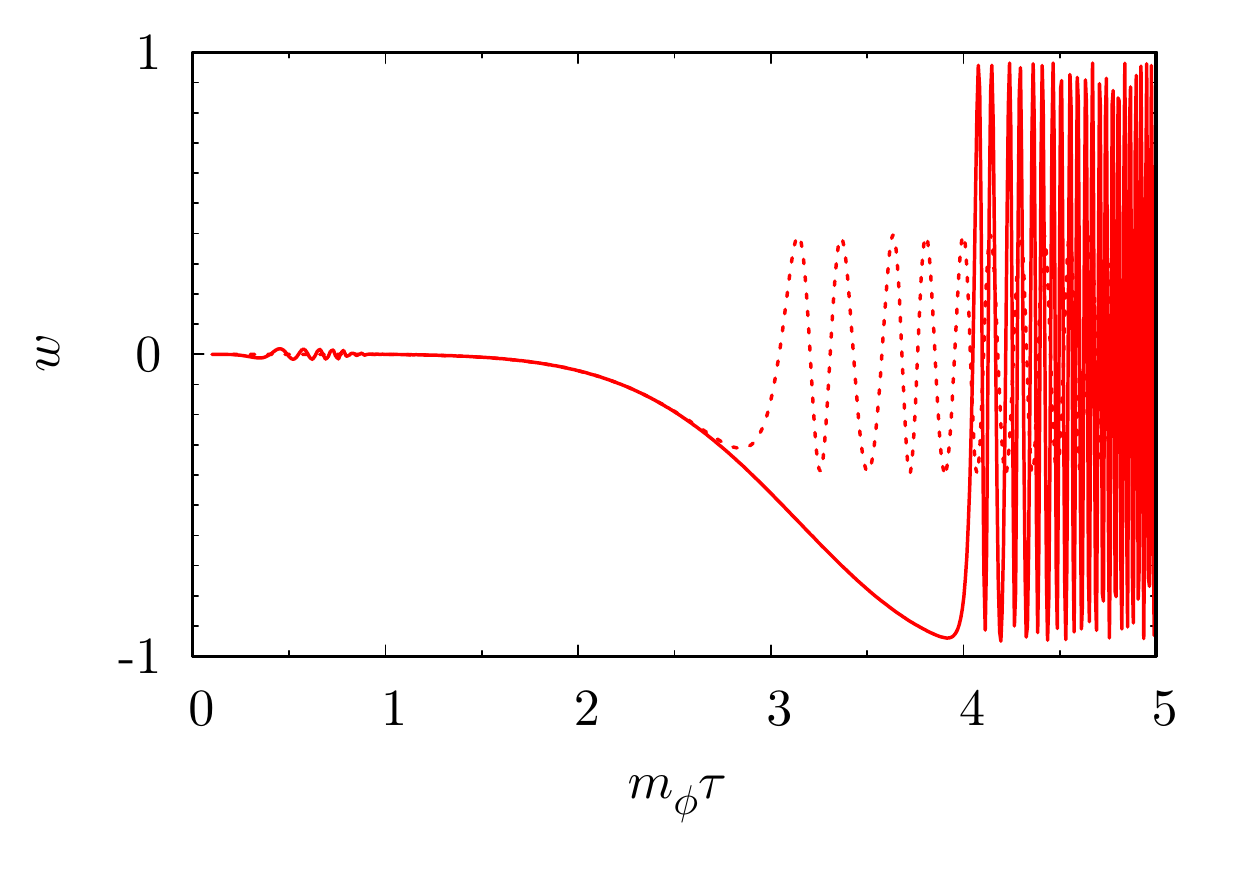}
\caption{
Left : time evolution of the energy density of the axion (red), dark photon field (green), Higgs (blue) and background matter (magenta) normalized by $m_\phi^2f_\phi^2$.
Right : time evolution of the equation of state parameter. The dotted line represents the case without dark photon production. $w$ approaches $-1$ when the dark Higgs is trapped at the origin.
}
\label{fig:rho_eos}
\end{figure*}

In Fig.~\ref{fig:field_higgsmass} we show the evolution of the spatially averaged field value (left) and the effective Higgs mass squared (right).  
One can see from the left panel  that the field value of the dark photon (green) becomes comparable to that of the axion (red) as a result of the tachyonic growth around $m_\phi \tau \simeq 1$. The dark Higgs field (blue) is stabilized at the origin by the effective mass determined by the field value of the dark photon, and it exhibits damped oscillations around the origin.
Then, at $m_\phi \tau \simeq 3$, the origin of the Higgs potential is destabilized when the field value of the dark photon becomes smaller than a critical value $\sim e^{-1} m_\Psi$ shown as the dashed horizontal line in the figure. In other words, as shown in the right panel, the effective Higgs mass squared from the dark photon field (green) becomes smaller than the negative mass squared of the Higgs potential (dashed magenta). Then, the dark Higgs develops a VEV and oscillates around it. Our numerical simulation explicitly shows that the destabilization can be significantly delayed due to the the non-thermal effective mass. For comparison, the dotted lines in the both panels show the case 
in which there is no dark photon production and the origin of the Higgs potential is destabilized earlier. 
Fig.~\ref{fig:rho_eos} shows the evolution of the energy density for each component (left) and the equation of state parameter (right).
Most importantly, the left panel shows that the Higgs vacuum energy density can  exceed the background matter density while the Higgs is trapped at the origin, and then, the trapped inflation occurs.
One can also find from the right panel that the equation of state parameter, $w$, becomes close to $-1$, showing a short period of inflation. 
Note that we have chosen the model parameters so that the inflation ends soon after it starts, because of the limited computational resource.

\begin{figure}[t!]
\includegraphics[width=8.5cm]{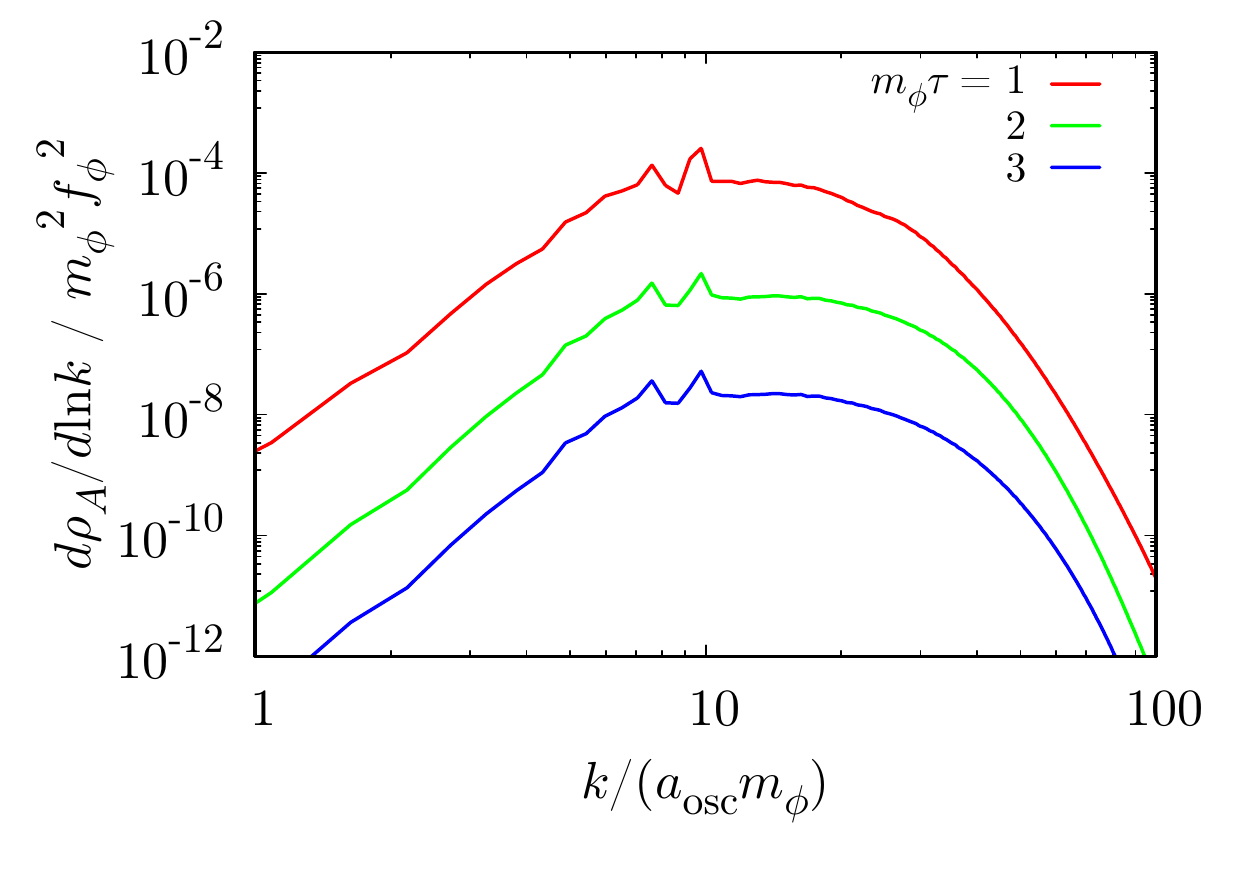}
\centering
\caption{
Spectrum of the energy density of the dark photon field, $\rho_A$, normalized by $m_\phi^2 f_\phi^2$, evaluated at $m_\phi\tau=1$ (red), 2 (green), and 3 (blue). The comoving wavenumber corresponding to the peak stays almost constant.}
\label{fig:spectrum}
\end{figure}

We show in Fig.~\ref{fig:spectrum} the spectrum of
the energy density of dark photons. One can see that
the momentum of the dark photon has a peak around
$k/a_{\rm osc} \simeq 10 m_\phi$, as expected from the
analytic estimate. The typical momentum is much smaller than the field value, and the peak momentum is more or less red-shifted by the cosmic expansion after the
system enters non-linear regime and the explosive production
stops. 

Although the above lattice simulation can only be performed within  finite parameter ranges due to the limited computational resource, we have confirmed that the produced dark photons indeed trap the dark Higgs at the origin for a while, and that inflation actually begins, and ends when
the effective mass becomes small.
This behavior was expected from the analysis in the previous subsection. Therefore, we believe that this numerical result supports that our analysis can also be applied to the case where the trapped inflation lasts longer.

\section{Cosmological moduli problem}
\label{sec:4}
In the previous section, we have shown that tachyonic production of dark photons can trap the dark Higgs at the origin, leading to inflation. In this section, we will study if the cosmological moduli problem can be solved (or mitigated) by the non-thermally trapped inflation.

From the analysis of Sec.~\ref{sec:duration}, the typical e-folding number of trapped inflation is estimated to be around $10 - 20$. In this section, we evaluate the extent to which the modulus abundance is diluted by the entropy production, and check whether the cosmological moduli problem can be solved. To be conservative, we will assume that the modulus $\chi$ dominated the universe before trapped inflation. If the modulus field only accounts for a fraction of the total energy, then it would become easier to solve the cosmological moduli problem by that amount.

The modulus abundance after the entropy production 
is given by\footnote{Here, 
we evaluate the modulus abundance assuming that it has not decayed until the entropy production.}
\beq
\frac{\rho_{\chi}}{s} = \frac{\rho_{\chi, {\rm{reh}}}}{s_{\rm{reh}}} \simeq \frac{3T_{\rm{reh}}}{4}e^{-3\mathcal{N}},
\eeq
where the subscript `reh' implies that
the variables are evaluated at the reheating, and we have assumed that
the dark Higgs behaves as non-relativistic matter after inflation until reheating. 
One of the features of the trapped inflation is that
the dark Higgs potential does not have to be extremely flat
unlike thermal inflation. Thus, the dark Higgs may soon decay
into lighter particles such as dark photons  after
inflation. In the case of such instantaneous reheating, the whole vacuum energy of Higgs is converted to the radiation energy, and the reheating temperature is given by
\beq
T_{\rm{reh}} &\simeq& \left(\frac{30 V_0}{\pi^2g_*(T_{\rm{reh}})}\right)^{1/4}\nonumber\\
&\simeq& 4\, {\rm GeV}\,\lambda^{-\frac{1}{4}}\left(\frac{100}{g_*(T_{\rm reh})}\right)^\frac{1}{4} \left(\frac{m_\Psi}{10\GeV}\right).
\eeq
We will see in the next section that such instantaneous reheating is
plausible in a broad parameter region.

The modulus abundance is constrained by observations of the light element abundances and  X-ray and gamma-ray backgrounds, depending on the mass and lifetime. 
For example,  the bound on the modulus abundance is approximately~\cite{Kawasaki:2017bqm} 
\beq
\frac{\rho_{\chi}}{s}
\lesssim 10^{-14} {\rm\,GeV}
\eeq
for $m_\chi\sim 1\TeV$ and the lifetime $\tau_\chi \gtrsim 10^4$\,sec.
Assuming the Planck-suppressed dimension five couplings to the
SM gauge bosons, the lifetime of the modulus is approximately $\tau_\chi \sim {\cal O}(10^4) (m_\chi/{\rm TeV})^{-3}$ sec. 
Thus, we have the lower bound on the e-folding number
\beq
\mathcal{N} \gtrsim 11+\frac{1}{3}\log\left(\frac{T_{\rm{reh}}}{1\GeV}\right)-\frac{1}{3}\log\left(
\frac{\rho_{\chi}/s}{10^{-14} \GeV}\right)\nonumber\\
\eeq
to satisfy the BBN bound on the modulus abundance. 
Since the typical value of $\mathcal{N}$ is about $10 - 20$ in our non-thermal trapped inflation, the moduli problem can be solved, or at least, greatly alleviated.

\begin{figure}[t!]
\includegraphics[width=8.5cm]{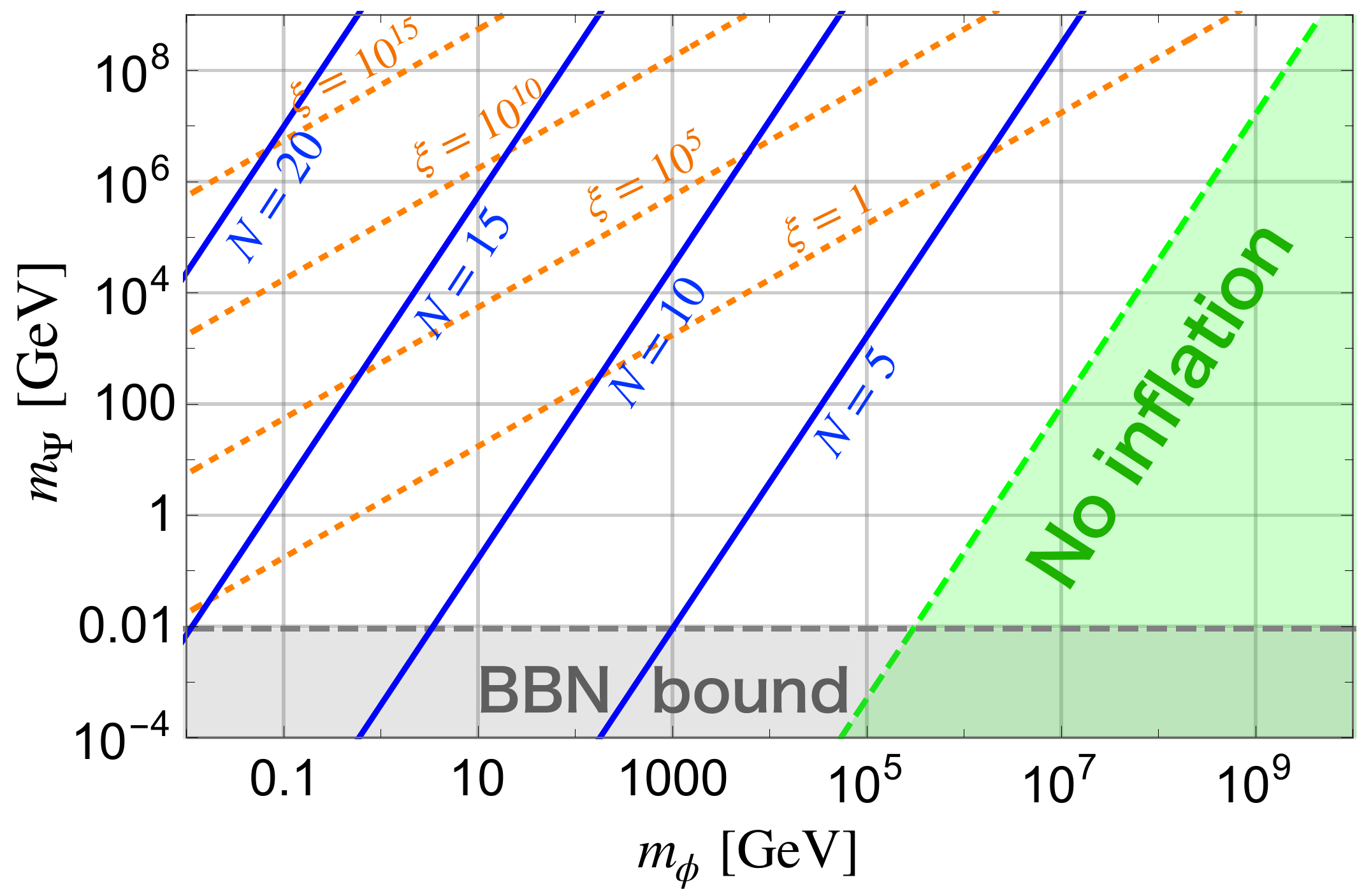}
\centering
\caption{
Contours of the e-folding number $\mathcal{N}$ are
shown by the blue solid lines in the $(m_\phi, m_\Psi)$ plane, where we take $\beta=30$, $e=0.1$, $\lambda=1$, and $f_\phi=10^{17}\GeV$. The gray shaded region denotes the BBN bound on the reheating temperature,
$T_{\rm reh}\gtrsim 4 \MeV$, and no trapped inflation in
the green shaded region.   The orange dotted lines represent the required values of $\xi$ to trap the Higgs before the onset of the axion oscillations, determined by $\xi R_{\rm osc}=m_\Psi^2$.
}
\label{fig:allowed}
\end{figure}

In Fig.~\ref{fig:allowed} we show the contours of 
the e-folding number $\mathcal{N}$ given by (\ref{efold}) on the $(m_\phi, m_\Psi)$ plane by blue solid lines. Here we take $\beta=30$, $e=0.1$, $\lambda=1$, and $f_\phi=10^{17}\GeV$. The gray shaded region denotes the BBN bound on the reheating temperature, $T_{\rm reh} \gtrsim 4 \MeV$~\cite{Kawasaki:1999na,Kawasaki:2000en,Ichikawa:2005vw}, where we assume the instantaneous reheating. In the lower right green region, the inflation does not occur, since dark photons produced by such heavy axion cannot keep the dark Higgs at its origin until the universe is dominated by the potential energy. The orange dotted contours show the required 
values of the non-minimal coupling $\xi$ to keep the Higgs at the origin until the onset of the axion oscillation, which is determined by $\xi R_{\rm osc}=m_\Psi^2$. 
In the region below $\xi=1$, we do not need to introduce the non-minimal coupling as long as the initial position of the Higgs field is sufficiently close to the origin. 
If we limit ourselves to the region below $\xi={\cal O}(1)$, the e-folding number ${\mathcal N}$ exceeds $10$ for $m_\phi \lesssim 100 \GeV$ and $m_\Psi \lesssim 100 \GeV$ for the adopted parameters.

\section{Reheating and experimental implications}
\label{sec:5}
In this section we study the reheating process of the dark Higgs after the trapped inflation. The origin  is destabilized and 
the dark Higgs field starts to develop a nonzero VEV when the effective mass due to the coupling to dark photons becomes smaller than $m_\Psi$. Let us expand the radial component of the Higgs field around the VEV as
\begin{align}
    |\Psi| = v + \frac{s}{\sqrt{2}},
\end{align}
where the $s$ field has mass $m_s = \sqrt{\lambda} v = \sqrt{2} m_\Psi$. The dark photon acquires a mass,
\begin{align}
    m_{\gamma'} = \sqrt{2} e v,
\end{align}
after the spontaneous symmetry breaking of U(1)$_H$.

After the inflation ends, the universe is dominated by the coherent oscillations of $s$. Then, if $m_s > 2 m_{\gamma'}$ i.e. $\lambda>8e^2$, the $s$ field quickly decays into a pair of dark photons. If the dark photon has a nonzero kinetic mixing $\epsilon$ with the SM photon (or hypercharge), it will further decay into the lighter SM fermions. Let us consider a dark photon decaying into a fermion anti-fermion pair.
The decay rate for $\gamma'\rightarrow f\bar{f}$ is
\beq
\Gamma_{\gamma'\rightarrow f\bar{f}}&=&\frac{N_c}{3}(\epsilon x_f)^2\alpha_{\rm{em}} m_{\gamma'}\sqrt{1-\frac{4m_f^2}{m_{\gamma'}^2}}\left(1+\frac{2m_f^2}{m_{\gamma'}^2}\right)\nonumber\\
&\equiv&\frac{1}{3}\epsilon^2\alpha_{\rm{em}}m_{\gamma'}\gamma_f(m_{\gamma'}),
\label{decay}
\eeq
where $\alpha_{\rm{em}}=e_{\rm{em}}^2/4\pi$ is the electromagnetic fine structure constant, $m_f$ is the fermion mass, $x_f$ is the magnitude of fermion charge (e.g. $x_u=2/3$ for up quark), and $N_c$ is the degree of freedom of color. $\gamma_f$ is defined as a function of $m_{\gamma'}$, which depends on the Yukawa coupling, electric charge, and color charge. Note that for $m_{\gamma'}\lesssim2\GeV$ we cannot estimate the hadronic decay simply by summing the $q\bar{q}$ contributions from \eq{decay}. The decay rate into hadrons is known to be obtained by \cite{Ilten:2018crw}
\beq
\Gamma_{\gamma' \rightarrow{\rm hadrons}}=\Gamma_{\gamma'\rightarrow\mu^+\mu^-}\mathcal{R}_\mu,
\eeq
where $\mathcal{R}_\mu\equiv\sigma(e^+e^-\rightarrow\rm{hadrons})/\sigma(e^+e^-\rightarrow\mu^+\mu^-)$ is determined by experiments. The contributions from charged leptons, charm quark, and bottom quark are taken into account by using (\ref{decay}).
Since $\Gamma_{s\rightarrow\gamma'\gamma'}/\Gamma_{\gamma'\rightarrow f\bar{f}}\propto(m_s/m_{\gamma'})^3\gg1$,  the reheating temperature is determined by the relation between the total decay rate $\Gamma_{\gamma'{\rm tot}}$ and the Hubble parameter at the end of inflation, $H_{\rm{end}}$.

If the kinetic mixing is sufficiently small, we have $\Gamma_{\gamma'{\rm tot}}<H_{\rm{end}}$, in which case the reheating does not complete until the Hubble parameter becomes equal to the decay rate of dark photon.
On the other hand, if $\Gamma_{\gamma'{\rm tot}}>H_{\rm{end}}$, the reheating is considered to be almost instantaneous. Here we neglect the Lorentz boost of the produced dark photons, assuming $e$ and $\lambda$ are of ${\cal O}(1)$.
Thus, the reheating temperature is given by 
\beq
\frac{T_{\rm{reh}}}{\GeV}\simeq
\begin{cases}
0.17\left(\frac{g_*(T_{\rm{reh}})}{40}\right)^{-1/4}\left(\frac{\epsilon}{10^{-8}}\right)\left(\frac{m_{\gamma'}}{100\MeV}\right)^{1/2}\vspace{1mm}\\
\hspace{30mm}(\Gamma_{\gamma'\rightarrow e^+e^-}<H_{\rm{end}})\\
\\
0.26\lambda^{1/4}\left(\frac{g_*(T_{\rm{reh}})}{40}\right)^{-1/4}\left(\frac{e}{0.1}\right)^{-1}\left(\frac{m_{\gamma'}}{100\MeV}\right),\vspace{1mm}\\
\hspace{30mm}(\Gamma_{\gamma'\rightarrow e^+e^-}>H_{\rm{end}})\nonumber
\end{cases}\\
\eeq
where in the first line only the decay into electrons is considered for the reference value of $m_{\gamma'}$, while for heavier $m_{\gamma'}$, contributions of decays into hadrons and the other leptons must be included.

On the other hand, for  $m_{\gamma'}<2m_e$, the decay into three photons is the leading one. However, the amplitude is suppressed by a fermion loop, which makes the reheating temperature too low, $T_{\rm{reh}}\simeq 200\KeV(\epsilon/0.1) (m_{\gamma'}/m_e)^9$. Thus, the successful reheating through dark photons is not possible if $m_{\gamma'}<2m_e$, and we need to assume that the dark Higgs mainly decays into the SM sector by other processes such as a portal coupling to the SM Higgs (see discussion below).

\begin{figure}[t!]
\includegraphics[width=8.5cm]{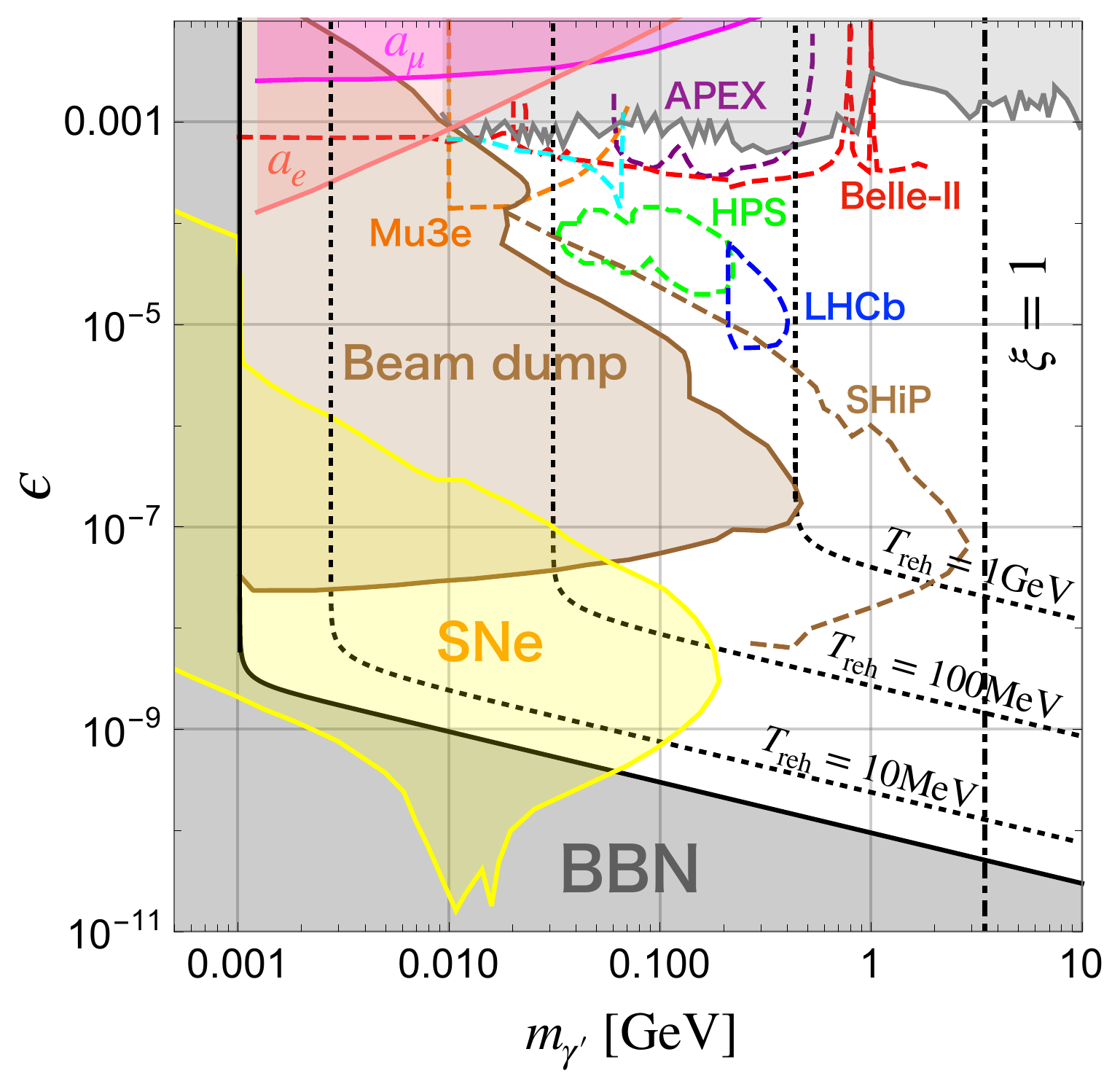}
\centering
\caption{The viable parameter region for the dark photon mass and the kinetic mixing, and the current constraints as well as projected sensitivities are shown. The shaded regions are excluded, and the colored dashed curves indicate expected sensitivity reach of future experiments. The dotted contours show the reheating temperature $T_{\rm reh} = 10\MeV, 100\MeV$, and $1\GeV$. The vertical dot-dashed line is an upper bound on $m_{\gamma'}$ required for trapping the dark Higgs before the axion oscillation when $\xi = 1$. See the text for details.
}
\label{fig:region}
\end{figure}

In \FIG{fig:region} we show the viable parameter space for the kinetic mixing as a function of the mass of dark photon, as well as the existing constraints and the future sensitivities. Here we take $e=0.1, \lambda=1$, and $m_\phi=10\GeV$. The black shaded region is excluded
by the constraint on the reheating temperature for the successful BBN, $T_{\rm reh}\gtrsim4\MeV$.
The colored shaded regions denote the existing bounds, and the dashed lines represent  the expected future sensitivity reach. The yellow one denoted by ``SNe" is the limit from the observation of SN1987A \cite{Chang:2016ntp}. The brown one denoted by ``Beam dump" represents constraints from the beam dump experiments including E141, E137, E774, KEK, Orsay, NA64, CHARM, $\nu$-Cal I and U70, see \cite{Andreas:2012mt} and references therein. The upper gray region represents the bound obtained by the direct mediator search in the visible decays of $\gamma'\rightarrow l^+l^-$, such as KLOE, NA48/2, HADES, PHENIX, A1, BaBar, and engineering runs for HPS and APEX, see \cite{Battaglieri:2017aum} and references therein. Limits from the anomalous magnetic moment of the electron and muon are also represented by the upper left pink and magenta regions, respectively \cite{Davoudiasl:2012ig, Endo:2012hp}. One can see that a broad parameter space region is expected to be probed by future experiments,
 as indicated by colored dashed curves: Belle-II (red), APEX (purple), MMAPS (cyan), HPS (green), LHCb (blue), and SHiP (brown), see \cite{Battaglieri:2017aum} and references therein. The dotted contours indicate the 
 reheating temperature, $T_{\rm reh} = 10\MeV, 100\MeV$ and $1 \GeV$.
 The black dot-dashed line denotes the required upper bound on $m_{\gamma'}$ to trap the dark Higgs before the onset of the axion oscillation, determined by $\xi R_{\rm osc}>m_\Psi^2$, where we show only the case of $\xi=1$. One can see from the figure that the kinetic mixing is bounded below for a given $m_{\gamma '}$ for successful reheating, and the kinetic mixing and mass have a one-to-one correspondence to the reheating temperature of the universe.

Even if the direct decay into dark photons is kinematically forbidden, $m_s<2m_{\gamma'}$ i.e. $\lambda<8e^2$, the decay of $s$ can still proceed through diagrams with off-shell dark photons. However, if the mass hierarchy is large, the decay
rate tends to be suppressed leading to too low reheating temperature. To complete the reheating in this case, we may consider some portal couplings. For example, we can consider
the Higgs portal coupling $\mathcal{L}\supset-\lambda_p|\Psi|^2 |H|^2$,
where $H$ denotes the SM Higgs doublet. 
The $s$ field with the mass $m_s>2m_H$ can decay into the SM Higgs pair, and for $m_s<2m_H$, the $s$ decays into the SM fermions through a mixing with the SM Higgs. As long as the mass of $s$ is heavier than twice the electron mass and the portal coupling is not suppressed, the reheating is considered to be instantaneous. 

We emphasize that, even in the presence of the Higgs portal coupling with $\lambda_p = {\cal O}(1)$, the $s$ mainly decays into the dark photons if kinematically allowed. This is because the latter is enhanced by the longitudinal modes. Thus, the above argument on the reheating process via the dark photon production is still valid in the presence of the Higgs portal coupling of ${\cal O}(1)$. On the other hand, if the decay into dark photons is not kinematically accessible, the decay through the Higgs portal coupling provides an alternative reheating process. 
In any case, the reheating is very efficient, and in fact, it is almost instantaneous for a large parameter space. This is  due to the fact that the trapped inflation does not require a very flat potential for the Higgs thanks to the intense trapping effect.

\section{Discussion and conclusions}
\label{sec:conclusion}
So far we have studied the trapped inflation and its phenomenological implications based on a simple low-energy effective theory including the dark Higgs field, dark photon, and axion. Here we discuss its possible UV completion in a SUSY framework.
One of the important requirement for the trapped inflation is the relatively light mass of the dark Higgs, and in particular, it must be lighter than the axion, $m_\Psi \lesssim m_\phi$, if the non-minimal coupling $\xi$ is
 of ${\cal O}(1)$.
For instance, such mass hierarchy can be understood if the hidden Higgs acquires a soft SUSY breaking mass only through Planck-suppressed interactions with the SUSY breaking sector, and if the axion (and the corresponding saxion) is stabilized {\it a la} KKLT~\cite{Kachru:2003aw}.
In this case we expect $m_\Psi$ is of order the gravitino mass $m_{3/2}$, and the axion mass is $m_\phi = {\cal O}(10-100)m_{3/2}$.   
In addition to the cosmological moduli problem, there are also the gravitino problem~\cite{Weinberg:1982zq,Krauss:1983ik} as well as the moduli-induced gravitino problem~\cite{Endo:2006zj,Nakamura:2006uc,Dine:2006ii,Endo:2006tf}, but these problems are also solved or ameliorated significantly in the presence the large entropy production. 

The entropy production dilutes not only the moduli but also any pre-existing baryon asymmetry and dark matter. Therefore, it is necessary to have a baryogenesis scenario that operates at low energy after the entropy production, or to create a very large baryon asymmetry beforehand. In the former case, electroweak baryogenesis, and in the latter case, the Affleck-Dine mechanism, are such candidates. One of the dark matter candidates  is stable moduli, which can explain dark matter if they dominate the universe before non-thermally trapped inflation starts. Another candidate is the QCD axion, which is generated around the QCD scale.

So far, we have focused our discussion on the scenario in which non-thermally trapped inflation leads to large entropy generation. Our setup may be applicable to other cosmological scenarios. For example, if the dark Higgs vacuum energy dominates the universe around the matter-radiation equality, it might behave as the early dark energy \cite{Poulin:2018cxd,Agrawal:2019lmo}. After a very short period of inflation, the dark Higgs will decay into dark photons, which will behave as dark radiation if the mass hierarchy is large enough. Alternatively, the dark photons may decay through kinetic mixing into massless fermions charged under another hidden U(1)$_{\rm H}'$, and these fermions may behave as dark radiation. Such early dark energy is motivated as a solution
to the Hubble tension, which is the discrepancy between the value of the Hubble constant $H_0$ inferred from the CMB observation \cite{Planck:2018vyg} and the standard cosmology and that from the observation of the late-time universe \cite{Riess:2016jrr} (for recent reviews, see \REF{Knox:2019rjx, Verde:2019ivm, DiValentino:2020zio}). The observational implications of applying non-thermally trapped inflation to early dark energy will be investigated in the future. It may be possible to test this scenario through gravitational waves~\cite{Adshead:2018doq,Machado:2018nqk,Adshead:2019lbr,Adshead:2019igv,Machado:2019xuc,Ratzinger:2020koh,Namba:2020kij,Kitajima:2020rpm,Okano:2020uyr,Geller:2021obo,Weiner:2020sxn}

Lastly let us comment on the possible importance of the Schwinger effect in our scenario. It is known that particle production occurs in a strong electric field, called the Schwinger effect \cite{Heisenberg:1936nmg,Schwinger:1951nm}. The dark photons are intensely produced in our model so that the Schwinger-like effect in the dark scalar QED may produce dark Higgs particles, which can severely hamper the tachyonic growth of dark photons. The production rate is suppressed by the exponential factor, $\exp(-\pi m_H^2/e|{\bm{E}}|)$, where $m_H^2=\xi R+e^2|\bm{A}|^2-m_\Psi^2$ and $|{\bm{E}}|~(\sim \omega_k|\bm{A}|)$ denotes the dark electric field. We note that this estimate here is somewhat rough because the suppression factor is applicable when the electric field is spatially uniform and constant with time. The production is most efficient when $e^2|\bm{A}|^2\sim \xi R$, since, after this time, the dark Higgs mass becomes heavier due to the produced dark photons. Thus, the Schwinger effect is considered to be suppressed if $\xi\gtrsim\beta^2/12\pi^2 = {\cal O}(1-10)$. Alternatively, one may assume that the dark Higgs has an extra mass through interactions with other moduli fields so that it remains sufficiently heavy during the exponential growth of the dark photon field.
On the other hand, if
 the dark Higgs initially has a vacuum expectation value, 
 the corresponding U(1)$_H$ is spontaneously broken and the dark photon is massive. As shown in Ref.~\cite{Agrawal:2018vin},
 the tachynic production of dark photons occurs if the dark photon mass is lighter than the axion mass. 
 As the  dark Higgs acquires an effective mass from dark photons, it gradually approaches the origin, and the U$(1)_H$ symmetry will be restored. In this process, the dark Higgs mass becomes lighter, and it becomes almost massless when $e|\bm{A}|\sim m_\Psi$. Then the dark Higgs may be copiously produced due to the Schwinger effect and the tachyonic production cannot proceed further.
A more precise estimate of the Schwinger effect in our scenario requires understanding of the production rate in a time-varying spatially non-uniform electric field, which is beyond the scope of this paper.

In this paper we have proposed a novel late-time inflation driven by the dark Higgs field, which is trapped at the origin through interactions with dark photons. Unlike thermal inflation, the momentum distribution of dark photons is significantly deviated from the thermal one, and it is dominated by low momentum modes produced by an axion condensate through tachyonic preheating. The trapping effect is much more intense than the thermal mass, which enables the dark Higgs to drive inflation even for a simple Mexican-hat potential. After the non-thermal inflation ends, the dark Higgs decays into massive dark photons, which further decay into the SM particles through a kinetic mixing. We have shown that a large portion of the viable parameter space can be probed by future experiments, partly because the kinetic mixing is bounded below for successful reheating. 

\vspace{2mm}
%
{\bf Acknowledgments.--}
We thank Prateek Agrawal and Matthew Reece for discussions and collaboration at an early stage. 
We also thank Masaki Yamada and Kazunori Nakayama for useful comments. 
FT thanks Harvard University for its hospitality where this work was initiated. 
The present work is supported by 
the Graduate Program on Physics for the Universe of Tohoku University (S.N.), JST SPRING, Grant Number JPMJSP2114 (S.N.),  Leading Young Researcher Overseas Visit Program at Tohoku University (F.T.)
JSPS KAKENHI Grant Numbers
17H02878 (F.T.), 19H01894 (N.K.), 20H01894 (F.T. and N.K.),
20H05851 (F.T. and N.K.), and 21H01078 (N.K.).
%

\bibliography{reference}

\end{document}